\documentclass[aps,plb, nofootinbib, notitlepage,onecolumn,superscriptaddress]{revtex4-1}
\usepackage{bm}
\usepackage{latexsym}
\usepackage{amsmath,amsfonts,amssymb}
\usepackage{graphicx,epsfig}
\usepackage{psfrag}
\usepackage{amsthm}
\usepackage{color}
\usepackage{braket}
\usepackage{indent first}
\usepackage[normalem]{ulem}
\usepackage{hyperref}
\usepackage[titletoc,title]{appendix}
 \usepackage{environ}
\usepackage{mathtools}
\usepackage{float}
\usepackage[makeroom]{cancel}
\usepackage{bbold}
\usepackage{footmisc}
\usepackage{titlesec}

\usepackage[utf8]{inputenc}
\newcommand{\be}{\begin{equation}}
\newcommand{\ee}{\end{equation}}
\newcommand{\bea}{\begin{eqnarray}}
\newcommand{\eea}{\end{eqnarray}}

\makeatletter
\newcommand\setcurrentname[1]{\def\@currentlabelname{#1}}
\makeatother

\NewEnviron{eqn}{
\begin{align}
\begin{split}
  \BODY
\end{split}
\end{align}
}

\NewEnviron{eqn*}{
\begin{align*}
\begin{split}
  \BODY
\end{split}
\end{align*}
}

\begin{document}
%%%%%%%%%%%%%%%%%%%%%%%%%%%%%%%%%%%%%%
\title{General Formalism of the Quantum Equivalence Principle}
%
%\title{The generalized Quantum Equivalence Principle}
%%%%%%%%%%%%%%%%%%%%%%%%%%%%%

\author{Saurya Das} 
\email{saurya.das@uleth.ca}
\affiliation{Theoretical Physics Group and Quantum Alberta, Department of Physics and Astronomy,
University of Lethbridge,
4401 University Drive, Lethbridge,
Alberta, T1K 3M4, Canada}

\author{Mitja Fridman}
\email{fridmanm@uleth.ca}
\affiliation{Theoretical Physics Group and Quantum Alberta, Department of Physics and Astronomy,
University of Lethbridge,
4401 University Drive, Lethbridge,
Alberta, T1K 3M4, Canada}

\author{Gaetano Lambiase} \email{lambiase@sa.infn.it}
\affiliation{Dipartimento di Fisica E.R: Caianiello, Universit\'a di Salerno, Via Giovanni Paolo II, 132 - 84084 Fisciano, Salerno, Italy}
\affiliation{INFN - Gruppo Collegato di Salerno, Via Giovanni Paolo II, 132 - 84084 Fisciano, Salerno, Italy}

%\affil[1]{Theoretical Physics Group and Quantum Alberta, Department of Physics and Astronomy,
%University of Lethbridge,
%4401 University Drive, Lethbridge,
%Alberta, T1K 3M4, Canada}
%\affil[2]{Dipartimento di Fisica E.R: Caianiello, Universit\'a di Salerno, Via Giovanni Paolo II, 132 - 84084 Fisciano, Salerno, Italy}
%\affil[3]{INFN - Gruppo Collegato di Salerno, Via Giovanni Paolo II, 132 - 84084 Fisciano, Salerno, Italy}

\begin{abstract}
\begin{center}
    \textbf{Abstract}\\
\end{center}
\par\noindent
%A consistent theory of quantum gravity will require a fully
%quantum formulation of the classical equivalence principle. 
%Recently, such a formulation in terms of the equality of the rest, inertial and gravitational mass operators, and for non-relativistic particles in a weak gravitational field was proposed by 
%Zych \& Brukner (2018). 
%
%In this work, the generalization to a fully relativistic formalism of the quantum equivalence principle,
%valid for all background space-times, as well as for massive bosons and fermions, is proposed 
%(the principle is trivially satisfied for massless particles). 
%It is shown that if the equivalence principle (i.e. the above operator equality)
%is broken at the quantum level, 
%it would imply the modification of the standard Lorentz transformations in flat space-time and 
%a corresponding modification of the metric in curved space-time by the different mass ratios. 
%In other words, the observed geometry would 
%effectively depend on the properties of the test particle. 
%Testable predictions of potential violations of the quantum equivalence principle are proposed.
%
A consistent theory of quantum gravity will require a fully quantum formulation of the classical equivalence principle. Such a formulation has been recently proposed in terms of the equality of the rest, inertial and gravitational mass operators, and for non-relativistic particles in a weak gravitational field. In this work, we propose a generalization to a fully relativistic formalism of the quantum equivalence principle, valid for all background space-times, as well as for massive bosons and fermions. The principle is trivially satisfied for massless particles. We show that if the equivalence principle is broken at the quantum level, it implies the modification of the standard Lorentz transformations in flat space-time and a corresponding modification of the metric in curved space-time by the different mass ratios. In other words, the observed geometry would effectively depend on the properties of the test particle.
%We propose examples of testable predictions of potential violations of the quantum equivalence principle.
Testable predictions of potential violations of the quantum equivalence principle are proposed.
\end{abstract}
\maketitle
%
%%%%%%%%%%%%%%%%%%%%%%%%%%%%%%%%%%%%%%%%%%%%%%%%%%%%%%
\section*{Introduction}
%%%%%%%%%%%%%%%%%%%%%%%%%%%%%%%%%%%%%%%%%%%%%%%%%%%%%%
%
%
%
%
The Equivalence Principle 
is the cornerstone of classical general relativity (GR)
\cite{Einstein,Weinberg:1972kfs,CMW}.
However, gravity is expected to be fundamentally a quantum force, just as the other three forces in Nature, with various approaches to quantum gravity being proposed \cite{qg1,Nicolai:2005mc,Amelino-Camelia:2010lsq}, it follows that there must exist a Quantum Equivalence Principle (QEP), from which the standard Equivalence Principle follows in the classical limit. 
Such a QEP was proposed by Zych \& Brukner \cite{Zych:2015fka}, which was applied to non-relativistic particles and weak gravitational fields. 
In this article, a necessary and sufficient
generalization of the applicability of the QEP to
relativistic bosons and fermions as well as to curved space-times is proposed. 
It is shown that if the QEP is violated, the Lorentz transformations in flat space-time and the corresponding metrics in curved space-times  
depend on the properties of the test particle itself and energies and momenta of non-interacting particles do not simply add up. In other words, if these problems are to be avoided, then the QEP should be exactly satisfied in Nature! 
Furthermore, not only should every theory of quantum gravity take into account this principle, but the latter may even provide useful guidance in the formulation of such theories. 

The QEP is proposed
in the form of an equality between the quantum 
operators representing 
the inertial, gravitational and rest masses of a given particle as \cite{Zych:2015fka}
\begin{eqnarray}
\hat{M}_\mathrm{I}=\hat{M}_\mathrm{G}=\hat{M}_\mathrm{R}~,
\label{qep10}
\end{eqnarray}
which in turn can be written in terms of the `ground state mass' $m_\alpha$ and the 
internal interactions of the composite particle, described by the internal energy operator $\hat{H}_{int,\alpha}$ as \cite{Zych:2015fka}
\begin{eqnarray}
\label{massop}
\hat{M}_\alpha=m_\alpha\hat{\mathbb{1}}+\frac{\hat{H}_{\mathrm{int,\alpha}}}{c^2}~,
\end{eqnarray}
where $\alpha=\mathrm{I,G,R}$ signify the inertial, gravitational and rest masses of the test particle,
and $c$ the speed of light. 
Eq. (\ref{qep10}) encompasses
all three features of the Equivalence Principle. $\hat{M}_\mathrm{I}=\hat{M}_\mathrm{G}$ represents the quantum version of the weak equivalence principle (WEP),
$\hat{H}_{\mathrm{int,R}}=\hat{H}_{\mathrm{int,I}}$ implies the quantum version of the local Lorentz invariance (LLI) and $\hat{H}_{\mathrm{int,R}}=\hat{H}_{\mathrm{int,G}}$ implies the quantum version of the local position invariance (LPI).
%Note however, that the formalism which led to the above relations is valid only for non-relativistic (i.e. slow-speed particles) and in a weak gravitational field that can be described by Newton's gravity.
Note that the formalism of Zych \& Brukner \cite{Zych:2015fka}, in which Eqs. (\ref{qep10}) and (\ref{massop}) are applied, is non-relativistic (i.e. slow-speed particles) and considers a weak gravitational field that can be described by Newton's gravity. In this article, it is shown 
%for the first time to the best of one's knowledge, 
that Eqs. (\ref{qep10}) and (\ref{massop}) can be applied in a generalized formalism, which can be used to test the QEP with massive bosons and fermions at relativistic speeds and in strong gravitational fields as well.

\section*{Results}
\label{res}

%To generalize the applicability of the above, one needs to consider modifications to special relativity and general relativity separately, and use relativistic quantum mechanics on curved space-time to arrive to a generalized formalism.

\subsection*{Modifications of Special Relativity}

First, as shown in Method \nameref{appSR}, when the three masses are not equal,
the standard dispersion relation of a free relativistic particle, namely
$E^2 = p^2c^2 + m^2c^4$, gets modified to the more general relation 
\begin{eqnarray}
\label{disprel}
E^2=\frac{m_\mathrm{R}}{m_\mathrm{I}}p^2c^2+m_\mathrm{R}^2c^4~,
\end{eqnarray}
which reduces to the dispersion relation, considered by Zych \& Brukner \cite{Zych:2015fka} in the non-relativistic limit and in the absence of a gravitational field.
This results in the following modified 
Lorentz transformation for any four vector, for a boost $\bf v$
\begin{eqnarray}
\label{lorentzt}
\Lambda_\nu^\mu(m_\mathrm{I}, m_\mathrm{G}, m_\mathrm{R})\equiv\Lambda_\nu^\mu(m_\alpha)=\left[\begin{array}{cccc}
\gamma     & -\gamma \sqrt{\frac{m_\mathrm{I}}{m_\mathrm{R}}}\frac{v_x}{c} & -\gamma \sqrt{\frac{m_\mathrm{I}}{m_\mathrm{R}}}\frac{v_y}{c} & -\gamma \sqrt{\frac{m_\mathrm{I}}{m_\mathrm{R}}}\frac{v_z}{c} \\
-\gamma \sqrt{\frac{m_\mathrm{I}}{m_\mathrm{R}}}\frac{v_x}{c}    & 1+(\gamma-1)\frac{v_x^2}{v^2} & (\gamma-1)\frac{v_xv_y}{v^2} & (\gamma-1)\frac{v_xv_z}{v^2} \\
-\gamma \sqrt{\frac{m_\mathrm{I}}{m_\mathrm{R}}}\frac{v_y}{c} & (\gamma-1)\frac{v_yv_x}{v^2} & 1+(\gamma-1)\frac{v_y^2}{v^2} & (\gamma-1)\frac{v_yv_z}{v^2} \\
-\gamma \sqrt{\frac{m_\mathrm{I}}{m_\mathrm{R}}}\frac{v_z}{c} & (\gamma-1)\frac{v_zv_x}{v^2} & (\gamma-1)\frac{v_zv_y}{v^2} & 1+(\gamma-1)\frac{v_z^2}{v^2}
\end{array}\right]~,
\end{eqnarray}
which is a transformation between reference frames with different velocities, e.g., $x^\mu=\Lambda_\nu^\mu(m_\alpha)\,x'^\nu$. The corrections in the $i0$ components appear as such, to preserve the norm $V_\mu V^\mu$ 
of an arbitrary 4-vector $V^\mu$  under boosts, and to be consistent with the modified Lorentz factor
\begin{eqnarray}
\label{gammaf}
\gamma=\frac{1}{\sqrt{1-\frac{m_\mathrm{I}}{m_\mathrm{R}}\frac{v^2}{c^2}}}~
\end{eqnarray}
as explained in Method \nameref{appSR}. Now, it follows from Eq.(\ref{lorentzt}) that 
%It turns out that
\begin{eqnarray}
\Lambda_\nu^\mu(m_{1,\alpha})p_1^\nu+\Lambda_\nu^\mu(m_{2,\alpha})p_2^\nu\neq\Lambda_\nu^\mu(m_{1,\alpha}+m_{2,\alpha})(p_1^\nu+p_2^\nu)~.
\end{eqnarray}
This means that even for two non-interacting particles, their masses do not just simply add up under the modified Lorentz transformations. 
This is known as the so-called soccer-ball problem \cite{Hossenfelder:2014ifa}, and the above result shows that if this problem is to be absent in Nature, the LLI must necessarily be satisfied. 
Note however, that regardless of whether LLI holds or is broken, transformation (\ref{lorentzt}) still forms a group. 
The generators of this group are still the standard Lorentz generators, since the modifications enter only through the boost parameters. 

%%%%%%%%%%START READING FROM HERE 16.01.2023
%\section{Applications}
%\label{APPS}

\subsection*{Modifications of General Relativity}

Next, one considers an arbitrary curved space-time, given by the (contravariant) metric $g^{\mu\nu}$.
For this, as shown in Method \nameref{appGR}, the metric $g^{\mu\nu}$ now contains 
%becomes a function of 
the ratios between inertial, gravitational and rest masses, i.e.,  $g^{\mu\nu}=g^{\mu\nu}(m_\mathrm{I},m_\mathrm{G},m_\mathrm{R})\equiv g^{\mu\nu}(m_\alpha)$. 
%As shows in the Appendix, this is   
%necessary and sufficient to obtain 
%
As expected, the standard form of the metric is recovered
when $m_\mathrm{I}=m_\mathrm{G}=m_\mathrm{R}$.
%
%Continue from here: 1.12.2022
%
%
%The formulation of the QEP must hold for any background space-time metric $g^{\mu\nu}$. 
%
To arrive at a specific background space-time, one must solve the Einstein equations
with the relevant stress-energy tensor and 
boundary conditions, 
and evaluate the integration constants, taking into account 
%while considering 
the masses $m_\alpha$. 
This is done by re-writing Newton's second law of motion and Newton's law of gravity for a freely falling particle with different inertial and gravitational masses as
\begin{eqnarray}
%\label{2ndNewt}
m_\mathrm{I}\,\mathbf{a}=-m_\mathrm{G}\,\nabla\phi\,\,\,\,\,\,\implies\,\,\,\,\,\,\frac{\mathrm{d}^2x^i}{\mathrm{d}t^2}=-\frac{m_\mathrm{G}}{m_\mathrm{I}}\partial_i\phi~,
\end{eqnarray}
where $\phi$ is the classical gravitational potential. A standard derivation of the metric from the Einstein equations follows, with the integration constants left arbitrary 
at first. To evaluate the integration constants, the weak field limit of the metric, where $g_{\mu\nu}=\eta_{\mu\nu}+h_{\mu\nu}$ and $|h_{\mu\nu}| \ll 1$ (here, $\eta_{\mu\nu}$ is the Minkowski flat space-time metric and $h_{\mu\nu}$ the small perturbation term), or equivalently, the weak field limit of the geodesic equation for that metric, is then compared to the above version of Newton's law of gravity. One must also simultaneously consider the 
modification of the relativistic energy-momentum
dispersion relation due to the different masses
%
%considering the modifications of special relativity 
(see Eq. (\ref{4velcomp}) in Method \nameref{appSR}). 
By doing so, the resultant metric obtains modification factors in the form of ratios $m_\alpha/m_{\alpha'}$, where $\alpha,\alpha'=\mathrm{I,G,R}$, as for example
\begin{eqnarray}
\label{weakfield}
    g_{00}=1+h_{00}=1+2\frac{m_\mathrm{G}}{m_\mathrm{R}}\frac{\phi}{c^2}~
\end{eqnarray}
for the {00} component of the weak gravitational field metric. For more general cases, where there is more than one integration constant, one must consider classical limits of other quantities, such as angular momentum of a rotating object in the case of the Kerr space-time (see Method \nameref{appK}), to obtain the modified integration constants.
This may have measurable consequences, as will be shown later. 
%In this section, three special cases of background space-times are considered, namely the space-time corresponding to a 
%spherically symmetric weak gravitational field, its generalization to the Schwarzschild space-time, and the Kerr space-time. 
%It turns out that such
Note that the mass ratios do not change the differential equations of GR, since they modify only the integration constants. 

Next, one considers the specific example of the 
Schwarzschild metric, 
which as one knows, is obtained by solving the vacuum Einstein equations under the assumptions of staticity, 
spherical symmetry and asymptotic flatness \cite{chandra}.
%Note that the metric does not depend on the azimuthal angle $\varphi$ of the spherical coordinate system. 
In the standard derivation of the Schwarzschild metric there is only one integration constant $K=-\frac{2GM}{c^2}$, where $G$ is the universal gravitational constant and $M$ the mass associated with the static, spherically symmetric metric. This constant is obtained in the asymptotic (weak field) limit, by comparing 
$|g_{00}-1|$ to the 
classical Newtonian potential $\phi=-\frac{GM}{r}$ (see Eq. (\ref{weakfield})), where $r$ is the distance from the central object. 
%
%where the field can be described by the classical Newtonian potential $\phi=-\frac{GM}{r}$. 
%
However, in the case where one distinguishes between the masses $m_\alpha$ of a test particle, it turns out (see Method \nameref{appW}) that these masses modify the integration constant $K=-\frac{m_\mathrm{G}}{m_\mathrm{R}}\frac{2GM}{c^2}$ and the metric reads as
\begin{eqnarray}
\label{schwarzmodm}
g^{\mu\nu}(m_\alpha)=\left[\begin{array}{cccc}
 \left(1-\frac{m_\mathrm{G}}{m_\mathrm{R}}\frac{2GM}{c^2r}\right)^{-1}    & 0 & 0 & 0 \\
0 & -\left(1-\frac{m_\mathrm{G}}{m_\mathrm{R}}\frac{2GM}{c^2r}\right) & 0 & 0 \\
0 & 0 & -\frac{1}{r^2} & 0 \\
0 & 0 & 0 & -\frac{1}{r^2\sin^2{\theta}}
\end{array}\right]~,
\end{eqnarray}
where $\theta$ the polar angle. 
In the above, one can see that the mass ratio $m_\mathrm{G}/m_\mathrm{R}$ modifies only the temporal and radial components of the metric (in the spherical coordinate system) and leaves the angular parts unchanged, while the inertial mass $m_\mathrm{I}$ does not affect the components in this case.
Therefore, if the ratio $m_\mathrm{G}/m_\mathrm{R}$ is universal, i.e., the same irrespective of the composition of the test particles, then it can be set to unity and the above metric becomes independent of the test masses, just as in standard GR. However, if the ratio does depend on the composition, then our result shows that it affects the effective space-time geometry itself, which should have observable consequences. 
Such effects can be studied in the weak gravitational field \cite{Schwartz:2019qqg}, or a general case of the Kerr space-time, where the modifications are obtained by following the procedure of Papapetrou \cite{Papapetrou:1974gq} (see Method \nameref{appK} for details) and reads as
\begin{eqnarray}
\label{kerrmod1}
g^{\mu\nu}(m_\alpha)=\left[\begin{array}{cccc}
 \frac{\Sigma^2(m_\alpha)}{\rho^2(m_\alpha)\Delta(m_\alpha)}    & 0 & 0& 2\frac{m_\mathrm{G}m_\mathrm{I}^{1/2}}{m_\mathrm{R}^{3/2}}\frac{GMar}{c^2\rho^2(m_\alpha)\Delta(m_\alpha)} \\
0 & -\frac{\Delta(m_\alpha)}{\rho^2(m_\alpha)} & 0 & 0 \\
0 & 0 & -\frac{1}{\rho^2(m_\alpha)} & 0 \\
2\frac{m_\mathrm{G}m_\mathrm{I}^{1/2}}{m_\mathrm{R}^{3/2}}\frac{GMar}{c^2\rho^2(m_\alpha)\Delta(m_\alpha)} & 0 & 0 & -\frac{\Delta(m_\alpha)-\frac{m_\mathrm{I}}{m_\mathrm{R}}a^2\sin^2{\theta}}{\rho^2(m_\alpha)\Delta(m_\alpha)\sin^2{\theta}}
\end{array}\right]~,
\end{eqnarray}
where $\Sigma^2(m_\alpha)=\big(r^2+\frac{m_\mathrm{I}}{m_\mathrm{R}}a^2\big)^2\!\!-\frac{m_\mathrm{I}}{m_\mathrm{R}}a^2\Delta(m_\alpha)\sin^2{\theta}$, $\,\,\Delta(m_\alpha)=r^2-\frac{m_\mathrm{G}}{m_\mathrm{R}}\frac{2GM}{c^2}r+\frac{m_\mathrm{I}}{m_\mathrm{R}}a^2$ and $\rho^2(m_\alpha)=r^2+\frac{m_\mathrm{I}}{m_\mathrm{R}}a^2\cos^2{\theta}$. From the above modified Kerr metric, one can see that all three masses $m_\alpha$ modify the metric in all coordinate directions, i.e., the temporal, radial, polar and azimuthal directions. If the rotation of the central object vanishes, i.e., $a\longrightarrow0$, the modified Schwarzschild solution from Eq. (\ref{schwarzmodm}) is recovered. Note that for the cases studied here, the inertial mass $m_\mathrm{I}$ enters the metric only in the rotating (non-static) case as seen from Eq. (\ref{kerrmod1}), where there is inertial frame dragging surrounding the central gravitating object. It turns out that $m_\mathrm{I}$ appears in the metric of any arbitrary rotating object, since in the asymptotic limit of the corresponding $g^{\mu\nu}$, one always obtains Eq. (\ref{asymlim}), where $m_\mathrm{I}$ enters through Eq. (\ref{modenmom}).

\subsection*{Generalized Formalism}
%\label{PROPS}

Equipped with the above building blocks, one now proceeds to
construct a formalism of the QEP, which is valid for bosons and fermions, travelling at arbitrary speeds and in arbitrary curved space-times. 
The corresponding wave equations reduce to
the standard Klein-Gordon and Dirac equations in curved space-time in the limit when the QEP is satisfied. 

First, one considers a bosonic particle, and writes its effective Hamiltonian in the Feshbach-Villars formalism  \cite{Feshbach:1958wv,Cognola:1985qg,Tagirov:1999nc}, 
suitably generalized to curved space-times and 
incorporating the three types of masses
%
%Once a distinction between the masses is made, it is straightforward to identify the rest and inertial masses in the Feshbach-Villars effective Hamiltonian on curved space-time
%
\begin{eqnarray}
\label{hamiltonianeq}
\hat{H}=\tau_3 \frac{1}{\sqrt{g^{00}(m_\alpha)}}m_\mathrm{R}c^2-(\tau_3+i\tau_2)\frac{g^{ij}(m_\alpha)}{\sqrt{g^{00}(m_\alpha)}}\frac{\hat{p}_i\hat{p}_j}{2m_\mathrm{I}}+(\tau_3+i\tau_2)\frac{g^{0i}(m_\alpha)}{\sqrt{g^{00}(m_\alpha)}}\frac{\hat{p}_0\hat{p}_i}{\sqrt{m_\mathrm{R}m_\mathrm{I}}}~,
\end{eqnarray}
where $\tau_k$ $(k=1,2,3)$ are the Pauli matrices. It can be easily shown that the above gives rise to 
the generalization of the 
Klein-Gordon equation in curved space-time \cite{qftcst} incorporating the three masses (see Method \nameref{appP}). 
%
%modifications. 
%The mass in the Klein-Gordon equation turns out as $m_\mathrm{R}$ and the covariant derivative obtains the modification $\sqrt{m_\mathrm{R}/m_\mathrm{I}}$ in the spatial part.
%
%to obtain a Hamiltonian in curved space-time for bosons and by using 
Note that in the above, the gravitational mass enters through the metric $g^{\mu\nu}(m_\alpha)$. It can also be straightforwardly shown that the above modified Hamiltonian exactly corresponds to the modified Klein-Gordon Lagrangian
\begin{eqnarray}
\mathcal{L}=\frac{1}{2}g^{\mu\nu}\!(m_\alpha)\,\tilde{\partial}_\mu\Phi_{_{\!\mathrm{KG}}}\tilde{\partial}_\nu\Phi_{_{\!\mathrm{KG}}}-\frac{1}{2}\frac{m_\mathrm{R}^2c^2}{\hbar^2}\Phi_{_{\!\mathrm{KG}}}^2~,
\end{eqnarray}
 where $\tilde{\partial}_\mu=\left(\frac{1}{c}\partial_0,\sqrt{\frac{m_\mathrm{R}}{m_\mathrm{I}}}\boldsymbol{\partial}\right)$. 
The above Lagrangian can be used to study the QEP with bosons in Quantum Field Theory in curved space-times. On variation, it 
gives rise to the same modified Klein-Gordon equation as obtained from 
Eq (\ref{hamiltonianeq}).
%and implicitly also through $\underline{\gamma^\mu}$ and $\Gamma_\mu$.

For fermions, on the other hand, one writes 
the standard Dirac Hamiltonian in curved space-time \cite{Parker:1980kw}, modified by the different masses as
\begin{eqnarray}
\label{dirachqep}
\hat{H}=\frac{1}{g^{00}(m_\alpha)}\sqrt{\frac{m_\mathrm{R}}{m_\mathrm{I}}}\underline{\gamma}^0(m_\alpha)\,\underline{\gamma}^i(m_\alpha)\,\hat{p}_i(m_\alpha)\,c+i\hbar\Gamma_0(m_\alpha)+\frac{1}{g^{00}(m_\alpha)}\,\underline{\gamma}^0(m_\alpha)\,m_\mathrm{R}c^2~,
\end{eqnarray}
where $\underline{\gamma}^\mu(m_\alpha)=e_a^\mu(m_\alpha)\,\gamma^a$ with $e_a^\mu(m_\alpha)$ being the vierbeins (or tetrads), defined by $g^{\mu\nu}(m_\alpha)=e_a^\mu(m_\alpha)\, e_b^\nu(m_\alpha)\,\eta^{ab}$, while $\gamma^a$ and $\eta^{ab}$ are the flat space-time Dirac gamma matrices and Minkowski metric, respectively, of the tangent space, whose basis is denoted by Latin indices. The derivative $\nabla_i(m_\alpha)$ from the momentum operator $\hat{p}_i(m_\alpha)=-i\hbar\nabla_i(m_\alpha)$ takes the form $\nabla_\mu(m_\alpha)=\partial_\mu-\Gamma_\mu(m_\alpha)$, where
\begin{eqnarray}
\Gamma_\mu(m_\alpha)=-\frac{1}{4}\gamma_a\gamma_b\,e_\nu^a(m_\alpha)\,g^{\nu\lambda}(m_\alpha)\,e_{\lambda;\mu}^b(m_\alpha)~
\end{eqnarray}
is the spinor affine connection. Note that $\underline{\gamma^\mu}(m_\alpha)$ and $\Gamma_\mu(m_\alpha)$ are functions of $m_\alpha$, since they depend on $g^{\mu\nu}(m_\alpha)$.  In the above $e_{\lambda;\mu}^b(m_\alpha)=\partial_\mu e_{\lambda}^b(m_\alpha)-\Gamma_{\mu\lambda}^\rho(m_\alpha)\, e_{\rho}^b(m_\alpha)$, where
\begin{eqnarray}
    \Gamma_{\mu\lambda}^\rho(m_\alpha)=\frac{1}{2}g^{\rho\sigma}(m_\alpha)\left(\partial_\mu g_{\lambda\sigma}(m_\alpha)+\partial_\lambda g_{\mu\sigma}(m_\alpha)-\partial_\sigma g_{\mu\lambda}(m_\alpha)\right)
\end{eqnarray} 
are the modified Christoffel symbols.

It can be easily shown that Eq. (\ref{dirachqep}) gives rise to the generalization of the Dirac equation in curved space-time incorporating  the three masses. It can also be shown that the above modified Hamiltonian corresponds exactly to the modified Dirac Lagrangian
\begin{eqnarray}
\mathcal{L}=\bar{\psi}\left(i\hbar \underline{\gamma}^\mu(m_\alpha)\,\tilde{\nabla}_\mu(m_\alpha)-m_\mathrm{R}c\right)\psi~,
\end{eqnarray}
where $\tilde{\nabla}_\mu(m_\alpha)=\left(\frac{1}{c}\nabla_0(m_\alpha),\sqrt{\frac{m_\mathrm{R}}{m_\mathrm{I}}}\,\boldsymbol{\nabla}(m_\alpha)\right)$. 
The above Lagrangian can be used to study the QEP with fermions in Quantum Field Theory in curved space-times, and gives rise to the same modified Dirac equation as obtained from Eq. (\ref{dirachqep}).
%Note that the second term acquires no $m_\alpha$ dependent 
%modification. 
%
In this case, the gravitational mass enters not only through the metric $g^{\mu\nu}(m_\alpha)$, but also implicitly through $\underline{\gamma^\mu}(m_\alpha)$ and $\Gamma_\mu(m_\alpha)$.

To describe the internal quantum effects of the test particle in the above generalized formalism, the masses $m_\alpha$ are promoted to quantum operators (see Eq. (\ref{massop})) in Eqs. (\ref{hamiltonianeq}) and (\ref{dirachqep}), which can be used to study the QEP in any given scenario.
%\begin{eqnarray}
%\label{massop}
%\hat{M}_\alpha=m_\alpha\hat{\mathbb{1}}+\frac{\hat{H}_{int,\alpha}}{c^2}~,
%\end{eqnarray}
Note that the operators $\hat{M}_\alpha$ describe the total mass-energy of a composite quantum system, with $m_\alpha$ being the ground state, and $\hat{H}_{\mathrm{int,\alpha}}$ effectively describe the relativistic internal effects of the composite particle, related to its degrees of freedom and drives the non-trivial internal evolution \cite{Zych:2015fka}.
This completes the 
generalization of the applicability of the QEP,
introduced Zych \& Brukner \cite{Zych:2015fka},
to relativistic bosons and fermions in arbitrary curved space-times.

\subsection*{Experimental Bounds}

The generalization of the applicability of the QEP proposed in this work may be tested in Earth-based experiments, such as interferometers and particle accelerators, and in strong gravitational fields, such as in the vicinity of stars and black holes. 
%There are several ways in which the QEP can be tested, 
The most common experiments are Earth-based experiments, which corresponds to a non-relativistic limit and a weak gravitational field. Such experiments include the neutron interferometer experiment by Colella, Overhauser and Werner (COW) \cite{Colella:1975dq} to test the WEP and other proposals \cite{Zych:2015fka} to test the LLI and the LPI. 
The COW experiment can be used to measure the effective gravitational mass $\langle\hat{M}_\mathrm{G}\rangle=m_\mathrm{G}+E_{\mathrm{int,G}}/c^2$ of the neutron, since the violation of the QEP modifies the phase shift to
\begin{eqnarray}
\Delta\Phi=\frac{m_\mathrm{G}\,g\,A}{\hbar\,v}+\frac{E_{\mathrm{int,G}}\,g\,A}{\hbar\,c^2v}~,
\end{eqnarray}
where $g$ is the gravitational acceleration, $A=lh$ is the area, $l$ the length and $h$ the height of the interferometer, $v$ is the velocity of the neutron in the lower branch and $E_{\mathrm{int,G}}=\langle H_{\mathrm{int,G}}\rangle$ is the deviation, obtained from the precision of measuring $\Delta\Phi$, which is $\sim10^{-3}$ \cite{Littrell:1997zz}. The effective inertial mass of the neutron corresponds to the accepted value of the mass of the neutron, since it is measured kinematically. Comparing the above masses provides an upper bound of $\lesssim10^{-3}$ for the deviation of $m_\mathrm{G}/m_\mathrm{I}$ from unity. The above bound corresponds to the upper bound on the difference of eigenvalues $\frac{1}{mc^2}(E_{\mathrm{int,G}}-E_{\mathrm{int,I}})$, where one assumes $m=m_\mathrm{G}=m_\mathrm{I}$ as the ground state value for both $\hat{M}_\mathrm{G}$ and $\hat{M}_\mathrm{I}$. Keeping the masses different would not change the order of magnitude of the correction. This provides a test for the WEP.

To test the QEP at relativistic speeds in the absence of a gravitational field, one can for example consider the mean lifetime $\tau$ of a particle in cosmic ray showers. It can easily be shown that the measured lifetime $\tau$ in a cosmic ray shower is given by
\begin{eqnarray}
\label{lifetime}
    \tau=\gamma\tau_0~,
\end{eqnarray}
where $\tau_0$ is the mean lifetime of the particle in its rest frame and $\gamma$ is defined in Eq. (\ref{gammaf}), containing the mass ratio $m_\mathrm{I}/m_\mathrm{R}$. Considering a charged pion decay, the mean lifetime in its rest frame is $\tau_0=26.0231\pm0.0050\,\mathrm{ns}$ \cite{Numao:1995qf}. It can be shown that this measurement provides\footnote{One can write the relevant mass ratio as 
\begin{eqnarray}
    \frac{m_\mathrm{I}}{m_\mathrm{R}}=1+\delta~, \nonumber
\end{eqnarray}
where $\delta\ll1$ is the deviation of $m_\mathrm{I}/m_\mathrm{R}$ from unity. By expanding $\gamma$ in Eq. (\ref{lifetime}) over $\delta$ and comparing the obtained deviation with the measurement uncertainty of $\tau$
\begin{eqnarray}
\sigma_\tau=\gamma\sigma_{\tau_0}\approx\gamma_0\sigma_{\tau_0}=\gamma_0^3\frac{v^2}{2c^2}\tau_0\,\delta~, \nonumber
\end{eqnarray}
where $\gamma_0=1/\sqrt{1-\frac{v^2}{c^2}}$ and $\sigma_{\tau_0}$ the uncertainty of $\tau_0$, one obtains the deviation in Eq. (\ref{LLIdev}). \label{foot1}}a speed dependent upper bound of 
\begin{eqnarray}
\label{LLIdev}
\delta\lesssim3.84\times10^{-4}\left(\frac{c^2}{v^2}-1\right)~
\end{eqnarray}
for the deviation of $m_\mathrm{I}/m_\mathrm{R}$ from unity. One can see that the upper bound becomes smaller as the speed of the pion increases. Pions with the highest measurable energies travel with speeds close to the speed of light $v=0.9999996c$ \cite{pions}. For such pions the above upper bound is $\delta\lesssim3.33\times10^{-10}$. The above bound corresponds to the upper bound on the difference of eigenvalues $\frac{1}{mc^2}(E_{\mathrm{int,I}}-E_{\mathrm{int,R}})$, where one assumes $m=m_\mathrm{I}=m_\mathrm{R}$ as the ground state value for both $\hat{M}_\mathrm{I}$ and $\hat{M}_\mathrm{R}$ for the same reason as in the WEP case. This provides a test for the LLI.

To test the QEP in strong gravitational fields and at relativistic speeds, one can for example consider the perihelion precession of planets. 
%Since this test is completely classical, there is no need to promote the masses to quantum operators. 
Following the standard procedure \cite{Weinberg:1972kfs} (p. 194-200) and using the metric in Eq. (\ref{schwarzmod}), it turns out that the perihelion precession during one orbit gets modified by the relevant mass ratio as
\begin{eqnarray}
\Delta\varphi=6\pi\frac{m_\mathrm{G}}{m_\mathrm{R}}\frac{GM}{c^2a(1-e^2)}~,
\end{eqnarray}
where $a$ is the semimajor axis and $e$ the eccentricity of the orbit. Considering a number of orbits $N$ around the central object, the above expression is multiplied by $N$ and the deviation should be detectable for a large enough $N$ if the EP is to be violated. It can be shown that the precision of the measured precession of the perihelion of Mercury $\Delta\varphi_{Mercury}=42.9799\pm0.0009\,\mathrm{\,"\,cy^{-1}}$ \cite{RSP} provides an upper bound of $\lesssim2.1\times10^{-5}$ for the deviation of $m_\mathrm{G}/m_\mathrm{R}$ from unity in a similar manner as seen in footnote \ref{foot1}. %A generalization for the Kerr metric in Eq. (\ref{kerrmod}) can be made as well to obtain higher order corrections.
The above bound corresponds to the upper bound on the difference of eigenvalues $\frac{1}{mc^2}(E_{\mathrm{int,G}}-E_{\mathrm{int,R}})$, where one assumes $m=m_\mathrm{G}=m_\mathrm{R}$ as the ground state value for both $\hat{M}_\mathrm{G}$ and $\hat{M}_\mathrm{R}$, for the same reason as in the WEP case. This provides a test for the LPI.

\section*{Conclusions}
%\label{CONC}

A precise formulation of the QEP is
essential for the formulation of quantum gravity, since 
gravity is an intrinsically quantum interaction in such a theory.  
%
%It gives insight in fundamental concepts which govern the QG energy scale.
In this work, adapting the formulation of Zych \& Brukner \cite{Zych:2015fka}, the most general formalism for the QEP is proposed, valid in generic curved space-times, for bosonic and fermionic particles, and moving with arbitrary velocities. 
The most important results that one obtains here are that the soccer-ball problem and LLI are related, and that the geometry, probed by a test particle, e.g., via the geodesic motion that it follows, in general depends on the ratios of the inertial, gravitational and rest masses of the test particle. In other words, the observed geometry is no longer just a function of the background metric, but also depends on the properties of the test particle (i.e., observer) itself, unless the QEP holds exactly! 
While it may be argued that such observer dependence of measurable physical quantities is already a feature of standard quantum mechanics, and that the above mass ratio dependence is present in our QEP, one notes here that the aforementioned test particle dependence should be there even at the purely classical gravity level. Although seemingly counter-intuitive, there is nothing intrinsically impermissible about it, and furthermore 
%one can even say that it is satisfying 
%it is interesting to note that 
such a necessary dependence from quantum mechanics now persists in its classical limit as well. 
%\newpage

%journal title, article title, volume number, page or article number or DOI, and year of publication

%%
%%  See where the following text fits in
%%
%%
%While the above general version of the QEP can be tested in astrophysical or terrestrial scenarios, as we shall see below, another remarkable fact comes out from our work. It is shown here that the geometry experienced by a test-particle, in general having different inertial, gravitational and rest masses $m_\mathrm{I},m_\mathrm{G}$ and $m_\mathrm{R}$, 
%in terms of its geodesics etc., depends not only on the background space-time, but also on the ratios of the above masses. This is true at the classical {\it and} quantum levels. 

\section*{Acknowledgement}

This work was supported by the Natural Sciences and Engineering Research Council of Canada. We thank the anonymous referees for their useful comments which helped in improving the manuscript.

\section*{Author contributions}

S. D., M.F. and G.L. have contributed to all aspects of the research, with the leading input from M.F.

\section*{Competing interests}

The authors declare no competing interests.

\section*{Data Availability}

Data sharing is not applicable to this work, since no data was used or generated in the process.

\section*{Methods}

\subsection*{Special Relativity}\setcurrentname{Special Relativity}
\label{appSR}

One first
considers the relativistic dispersion relation and its non-relativistic limit. From the non-relativistic limit of the energy-momentum dispersion relation one finds
\begin{eqnarray}
\label{classlimdr}
E=\sqrt{p^2c^2+m^2c^4}\approx mc^2+\frac{p^2}{2m}\,\,\,\,\,\,\longrightarrow \,\,\,\,\,\, m_\mathrm{R}c^2+\frac{p^2}{2m_\mathrm{I}}~,
\end{eqnarray}
where one distinguishes between the inertial and rest masses. One can trace back the steps, and in a straightforward manner identify where masses $m_\alpha$ enter the standard relativistic energy-momentum dispersion relation, which then reads
\begin{eqnarray}
\label{disprel}
E^2=\frac{m_\mathrm{R}}{m_\mathrm{I}}p^2c^2+m_\mathrm{R}^2c^4~.
\end{eqnarray}
Note that the form of Eq. (\ref{disprel}) is necessary to obtain the non-relativistic limit of Eq. (\ref{classlimdr}). 
Since there is no corresponding 
non-relativistic relation for massless particles, there is no equivalent of 
Eq. (\ref{disprel}) for photons.
Therefore, the issues related to the equivalence principle and its quantum
counterpart, as studied in this work, do not apply to massless particles.
To satisfy Eq. (\ref{disprel}) and the Lorentz scalar for the four-momenta $p^\mu p_\mu=m_\mathrm{R}^2c^2$, the four-momentum must take the form of
\begin{eqnarray}
\label{fourmom}
p^\mu=\left(\frac{E}{c},\sqrt{\frac{m_\mathrm{R}}{m_\mathrm{I}}}\mathbf{p}\right)~,
\end{eqnarray}
where $\mathbf{p}=m_\mathrm{I}\mathbf{v} \gamma$ is the relativistic momentum, $\gamma$ is the Lorentz factor and $E=m_\mathrm{R}c^2\gamma$ is the energy of the particle, equivalent to Eq. (\ref{disprel}). By using the definitions of $E$ and $\mathbf{p}$ containing the Lorentz factor $\gamma$, and plugging them in the Lorentz scalar product for four-momenta $p^\mu p_\mu=m_\mathrm{R}^2c^2$, one can immediately see that the Lorentz factor $\gamma$ also contains a ratio of inertial and rest masses
\begin{eqnarray}
\label{gammaf}
\gamma=\frac{1}{\sqrt{1-\frac{m_\mathrm{I}}{m_\mathrm{R}}\frac{v^2}{c^2}}}~.
\end{eqnarray}
The above modified Lorentz factor can be interpreted in terms of a modification to the time dilation and length contraction
of a relativistic particle. Note that the above Lorentz factor is also a function of $m_\alpha$. Specifically, a function of the ratio $m_\mathrm{I}/m_\mathrm{R}$. This is the case for all of the following considerations.

To define the four-velocity, one can write the four-momentum from Eq. (\ref{fourmom}) in terms of the four-velocity as
\begin{eqnarray}
\label{fourmomod}
p^\mu=\sqrt{m_\mathrm{R}m_\mathrm{I}}\,c\left(\sqrt{\frac{m_\mathrm{R}}{m_\mathrm{I}}}\gamma,\frac{\mathbf{v}}{c}\gamma\right)=\sqrt{m_\mathrm{R}m_\mathrm{I}}\,c\left(\frac{1}{c}\frac{c\,\mathrm{d}t}{\mathrm{d}\tau},\sqrt{\frac{m_\mathrm{I}}{m_\mathrm{R}}}\frac{1}{c}\frac{\mathrm{d}\mathbf{x}}{\mathrm{d}\tau}\right)=\sqrt{m_\mathrm{R}m_\mathrm{I}}\,c\,u^\mu~,
\end{eqnarray}
where factoring out the masses appears as such that $u^\mu u_\mu=\frac{m_\mathrm{R}}{m_\mathrm{I}}$ is satisfied. From the above, one can read out the zeroth component of the four-velocity as
\begin{eqnarray}
\label{4velcomp}
\frac{1}{c}\frac{c\,\mathrm{d}t}{\mathrm{d}\tau}=\sqrt{\frac{m_\mathrm{R}}{m_\mathrm{I}}}\gamma\,\,\,\,\,\,\implies\,\,\,\,\,\,\left(\frac{c\,\mathrm{d}t}{\mathrm{d}\tau}\right)^2=\frac{m_\mathrm{R}}{m_\mathrm{I}}c^2\gamma^2~.
\end{eqnarray}

Using the modified Lorentz transformation from the main paper with the four-velocity defined above, for a particle moving in the $x$-direction, one finds the composition law for velocities
\begin{subequations}
\begin{eqnarray}
\gamma'\!\!&=&\!\!\gamma\,\gamma_0\left(1-\frac{m_\mathrm{I}}{m_\mathrm{R}}\frac{v_0v}{c^2}\right)~, \\
\label{velb}
\mathrm{or}\,\,\,\,\,\,\gamma'\!\!&=&\!\!\gamma\,\gamma_0-\sqrt{\gamma_{\vphantom{0}}^2-1}\sqrt{\gamma_0^2-1}~, \\
\label{velc}
\implies\,\,\,v'\!\!&=&\!\!\frac{v_0-v}{1-\frac{m_\mathrm{I}}{m_\mathrm{R}}\frac{v_0v}{c^2}}~,
\end{eqnarray}
\end{subequations}
where $v_0$ is the particle velocity corresponding to $\gamma_0$ in the initial inertial frame, $v$ is the velocity of the boosted inertial frame corresponding to $\gamma$ and $v'$ is the particle velocity corresponding to $\gamma'$ as measured in the boosted inertial frame. It is interesting to note that Eq. (\ref{velb}) does not explicitly depend on $m_\alpha$, and is identical as in the standard case. However, the addition of velocities does depend on $m_\alpha$, as seen in Eq. (\ref{velc}), due to the dependence of the Lorentz factors on $m_\alpha$.

\subsection*{General Relativity}\setcurrentname{General Relativity}
\label{appGR}

Here, the formulation of the equivalence principle is generalized to curved space-times, using masses $m_\alpha$.

\subsubsection*{Weak Gravitational Field and the Schwarzschild Metric}\setcurrentname{Weak Gravitational Field and the Schwarzschild Metric}
\label{appW}
 
%
%To generalize the formulation of the equivalence principle in curved space-times, using distinguished masses, one needs to verify if such test masses enter the metric tensor in any way. 
%For the weak gravitational field limit, the test masses enter the metric in a similar way than for the Schwarzshild metric. 
%To see this, one takes the following approach.
In GR, the motion of a particle in a background space-time is described by the geodesic equation \cite{chandra}
\begin{eqnarray}
\label{geodesic}
\frac{\mathrm{d}^2x^\mu}{\mathrm{d}\tau^2}+\Gamma_{\rho\sigma}^\mu\frac{\mathrm{d}x^\rho}{\mathrm{d}\tau}\frac{\mathrm{d}x^\sigma}{\mathrm{d}\tau}=0~,
\end{eqnarray}
where $\tau$ represents the proper time. 
%affine parameter of motion.
In the case of the weak field limit of GR, when $g_{\mu\nu}=\eta_{\mu\nu}+h_{\mu\nu}$ and $|h_{\mu\nu}| \ll 1$ (in the following, all terms of second order and higher in $h_{\mu\nu}$ are ignored), and the Newtonian limit for which 
$\frac{\mathrm{d}x^i}{\mathrm{d}\tau}\ll\frac{c\,\mathrm{d}t}{\mathrm{d}\tau}$,
one has 
$\Gamma_{00}^i=-\frac{1}{2}\partial_ih_{00}$
and the geodesic equation reduces to
%
%one obtains the expression for the second time derivative of position from the geodesic equation to evaluate the 00 component of the metric. For this, one needs to consider a slowly moving particle $\frac{\mathrm{d}x^i}{\mathrm{d}\tau}\ll\frac{\mathrm{d}t}{\mathrm{d}\tau}$ (to compare to the Newtonian limit) and $\Gamma_{00}^i=-\frac{1}{2}\partial_ih_{00}$. The geodesic equation then reduces to 
%
\begin{eqnarray}
\label{geoapprox}
\frac{\mathrm{d}^2x^i}{\mathrm{d}\tau^2}\approx\frac{\mathrm{d}^2x^i}{\mathrm{d}t^2}=-\frac{1}{2}\partial_ih_{00}\left(\frac{c\mathrm{d}t}{\mathrm{d}\tau}\right)^2~.
\end{eqnarray}
To obtain $h_{00}$, one must identify $\frac{\mathrm{d}^2x^i}{\mathrm{d}t^2}$ and $\frac{c\,\mathrm{d}t}{\mathrm{d}\tau}$ in the non-relativistic limit, using test particle masses $m_\alpha$. For the former, one considers Newton's second law applied to gravity, with the inertial and gravitational masses assumed to be different
\begin{eqnarray}
\label{2ndNewt}
m_\mathrm{I}\,\mathbf{a}=-m_\mathrm{G}\nabla\phi~,
\end{eqnarray}
where $\phi$ is the gravitational potential and $\mathbf{g}=-\nabla\phi$ is the gravitational field. The acceleration can be written in component form as 
%the second derivative of position
%
\begin{eqnarray}
\label{newt2}
\frac{\mathrm{d}^2x^i}{\mathrm{d}t^2}=-\frac{m_\mathrm{G}}{m_\mathrm{I}}\partial_i\phi~.
\end{eqnarray}

Since one must consider the non-relativistic limit to obtain $h_{00}$, in which case we can set $\gamma=1$, since $v^2\ll c^2$.
To evaluate the small deviation (in this case of the 00 component) of the metric from the flat Minkowski space-time, due to a weak gravitational field, one plugs Eqs. (\ref{4velcomp}) and (\ref{newt2}) in Eq. (\ref{geoapprox}) to obtain
\begin{eqnarray}
\label{g00m}
h_{00}=2\frac{m_\mathrm{G}}{m_\mathrm{R}}\frac{\phi}{c^2}\,\,\,\,\,\,\implies\,\,\,\,\,\,g_{00}=1+h_{00}=1+2\frac{m_\mathrm{G}}{m_\mathrm{R}}\frac{\phi}{c^2}~.
\end{eqnarray}
Using the same approach, one obtains the modifications for other components of the metric. Because the equations in this work make use of a contravariant metric $g^{\mu\nu}$, one must calculate the inverse of $g_{\mu\nu}$. For the case of a diagonal metric, the inverse simply inverts the diagonal elements. Therefore,
\begin{eqnarray}
g^{00}=\frac{1}{g_{00}}=\frac{1}{1+2\frac{m_\mathrm{G}}{m_\mathrm{R}}\frac{\phi}{c^2}}\approx 1-2\frac{m_\mathrm{G}}{m_\mathrm{R}}\frac{\phi}{c^2}~,
\end{eqnarray}
since $\phi\ll c^2$ in the weak field limit. These results can then be used to write the weak field metric as
\begin{eqnarray}
\label{metricequiv}
g^{\mu\nu}=\left[\begin{array}{cccc}
1-2\frac{m_\mathrm{G}\phi}{m_\mathrm{R}c^2}  & \mathcal{O}(c^{-5}) \\
 \mathcal{O}(c^{-5})& -\mathbb{1}_{3x3}\left(1+2\frac{m_\mathrm{G}\phi}{m_\mathrm{R}c^2}\right) 
 %&  & -1-2\frac{\phi}{c^2} &  \\
%\mathcal{O}(c^{-5}) &  &  & -1-2\frac{\phi}{c^2}
\end{array}\right]~,
\end{eqnarray}
where one can see that the mass ratio $m_\mathrm{G}/m_\mathrm{R}$ modifies the metric in the temporal and all spatial components, while the inertial mass $m_\mathrm{I}$ does not affect it. The off-diagonal elements are of order $\mathcal{O}(c^{-5})$ and can be ignored. Note that the above metric is given in a Cartesian coordinate system for convenience of use in Earth based experiments.
It turns out that every $p^2$ obtains a factor $1/m_\mathrm{I}$ in front (or equivalently, every $v^2$ obtains a $m_\mathrm{I}$ in front), every $\phi$ obtains a factor $m_\mathrm{G}$ in front and every $c^2$ obtains a factor of $m_\mathrm{R}$ in front, which can be used as a rule of thumb to enter masses $m_\alpha$ in expressions.

%\section{Schwarzschild metric}
%\label{appS}

For a strong gravitational field, described by a Schwarzschild space-time, the general solution for a covariant metric, before evaluating the integration constant $K$, has the form
\begin{eqnarray}
\label{schwarzconst}
g_{\mu\nu}=\left[\begin{array}{cccc}
\left(1+\frac{K}{r}\right)  & 0    & 0  & 0 \\ 
0  &   -\left(1+\frac{K}{r}\right)^{-1}  & 0  & 0 \\
0  &    0 & -{r^2}  &  0\\
0  &    0 & 0  & -{r^2\sin^2{\theta}}
\end{array}\right]~.
\end{eqnarray}
The constant $K$ is obtained in the asymptotic limit, where $r\longrightarrow\infty$ and the gravitational field is weak. Therefore, one can use the result from Eq. (\ref{g00m}) to compare to the $g_{00}$ component of the above Schwarzschild metric. It follows that 
\begin{eqnarray}
\label{constK}
K=2\frac{m_\mathrm{G}}{m_\mathrm{R}}\frac{\phi}{c^2}r=-\frac{m_\mathrm{G}}{m_\mathrm{R}}\frac{2GM}{c^2}~.
\end{eqnarray}
The inverse of the metric from Eq. (\ref{schwarzconst}), using the above constant, exactly corresponds to  the modified Schwarzschild metric 
\begin{eqnarray}
\label{schwarzmod}
g^{\mu\nu}(m_\alpha)=\left[\begin{array}{cccc}
 \left(1-\frac{m_\mathrm{G}}{m_\mathrm{R}}\frac{2GM}{c^2r}\right)^{-1}    & 0 & 0 & 0 \\
0 & -\left(1-\frac{m_\mathrm{G}}{m_\mathrm{R}}\frac{2GM}{c^2r}\right) & 0 & 0 \\
0 & 0 & -\frac{1}{r^2} & 0 \\
0 & 0 & 0 & -\frac{1}{r^2\sin^2{\theta}}
\end{array}\right]~.
\end{eqnarray}
In the above one can see that the mass ratio $m_\mathrm{G}/m_\mathrm{R}$ modifies the metric only in the temporal and radial components, while the inertial mass $m_\mathrm{I}$ does not affect it. 
Note that in the weak field limit of the geodesic equation from Eq. (\ref{geodesic}) for the above metric, one obtains the second Newton's law from Eq. (\ref{2ndNewt}).

\subsubsection*{Kerr Metric}\setcurrentname{Kerr Metric}
\label{appK}

To obtain the additional integration constant $L$, due to the rotation of the central object, the procedure of Papapetrou \cite{Papapetrou:1974gq} is followed. 
The general form of a Kerr metric with the 
constants $K$ and $L$ can be written as \cite{chandra}
\begin{eqnarray}
\label{kerrconst}
g_{\mu\nu}=\left[\begin{array}{cccc}
 \left(1+\frac{Kr}{\rho^2}\right)    & 0 & 0& -\frac{KL\,r}{\rho^2}\sin^2{\theta} \\
0 & -\frac{\rho^2}{\Delta} & 0 & 0 \\
0 & 0 & -{\rho^2} & 0 \\
-\frac{KL\,r}{\rho^2}\sin^2{\theta} & 0 & 0 & -\left(r^2+L^2-\frac{KL^2r\sin^2{\theta}}{\rho^2}\right)\sin^2{\theta}
\end{array}\right]~,
\end{eqnarray}
where $\Delta=r^2+Kr+L^2$ and $\rho^2=r^2+L^2\cos^2{\theta}$. Since the procedure of Papapetrou \cite{Papapetrou:1974gq} makes use of Cartesian coordinates, it is convenient to write the above metric in Cartesian coordinates as a line element \cite{chandra}
\begin{eqnarray}
\label{kerrcart}
\mathrm{d}s^2&=&\left(1+\frac{Kr^3}{r^4+L^2z^2}\right)c^2\mathrm{d}t^2-\left(1-\frac{Kr^3}{(r^4+L^2z^2)(r^2+L^2)^2}(Ly-rx)^2\right)\mathrm{d}x^2 \nonumber \\
&-&\left(1-\frac{Kr^3}{(r^4+L^2z^2)(r^2+L^2)^2}(Lx+ry)^2\right)\mathrm{d}y^2-\left(1-\frac{Krz^2}{r^4+L^2z^2}\right)\mathrm{d}z^2 \nonumber \\
&+&\frac{2Kr^3}{(r^4+L^2z^2)(r^2+L^2)}(Ly-rx)\,\,c\,\mathrm{d}t\,\,\mathrm{d}x-\frac{2Kr^3}{(r^4+L^2z^2)(r^2+L^2)}(Lx+ry)\,\,c\,\mathrm{d}t\,\,\mathrm{d}y \nonumber \\
&-&\frac{2Kr^2z}{r^4+L^2z^2}\,\,c\,\mathrm{d}t\,\,\mathrm{d}z-\frac{2Kr^3}{(r^4+L^2z^2)(r^2+L^2)^2}(Ly-rx)(Lx+ry)\,\mathrm{d}x\,\,\mathrm{d}y \nonumber \\
&-&\frac{2Kr^2z}{(r^4+L^2z^2)(r^2+L^2)}(Ly-rx)\,\mathrm{d}x\,\,\mathrm{d}z+\frac{2Kr^2z}{(r^4+L^2z^2)(r^2+L^2)}(Lx+ry)\,\mathrm{d}y\,\,\mathrm{d}z~,
\end{eqnarray}
where $x=(r\cos{\varphi}+L\sin{\varphi})\sin{\theta}$, $y=(r\sin{\varphi}-L\cos{\varphi})\sin{\theta}$, $z=r\cos{\theta}$ and $r^4-r^2(x^2+y^2+z^2-L^2)-L^2z^2=0$. 

%Following the procedure outlined in Ref. \cite{Papapetrou:1974gq}, 
One obtains the metric of a rotating object in the asymptotic limit from the weak field Einstein equations ($g_{\mu\nu}=\eta_{\mu\nu}+h_{\mu\nu}$ with $|h_{\mu\nu}|\ll1$), which is later compared to the asymptotic limit of the metric in Eq. (\ref{kerrcart}), to obtain constants $K$ and $L$. The weak field Einstein equations read as \cite{Papapetrou:1974gq}
\begin{eqnarray}
\label{lineinst}
\Box\gamma_{\mu\nu}=-2\kappa \,\,_0T_{\mu\nu}~,
\end{eqnarray}
where 
\begin{eqnarray}
\label{gammah}
\gamma_{\mu\nu}=h_{\mu\nu}-\frac{1}{2}\eta_{\mu\nu}h~,\,\,\,\,\,\,h=\eta^{\rho\sigma}h_{\rho\sigma}~,
\end{eqnarray}
$\kappa$ is a constant, which takes the value $8\pi G/c^4$ in the standard case (for now $\kappa$ is taken to be arbitrary, because it obtains modifications, as seen in Eq. (\ref{einconst})) and $_0T_{\mu\nu}$ is the special-relativistic energy-momentum tensor which describes the source of the gravitational field for a weak gravitational field. The Kerr metric is stationary, which means that the temporal derivative in Eq. (\ref{lineinst}) vanishes and one obtains
\begin{eqnarray}
\label{statlineinst}
\nabla^2\gamma_{\mu\nu}=2\kappa \,\,_0T_{\mu\nu}~.
\end{eqnarray}
The general solution to the above equation reads as 
\begin{eqnarray}
\label{gammasol}
\gamma_{\mu\nu}(\mathbf{x})=-\frac{\kappa}{2\pi}\int\frac{1}{R}\,_0T_{\mu\nu}(\mathbf{X})\,\mathrm{d}^3\mathbf{X}~,
\end{eqnarray}
where $\mathbf{x}$ is an arbitrary location outside the gravitating object, $\mathbf{X}$ is the location inside the gravitating object, over which the integral is evaluated, and $R$ is the distance between the two locations, defined as
\begin{eqnarray}
R^2=(x^i-X^i)^2=r^2-2x^iX^i+X^iX^i~,
\end{eqnarray}
where the Einstein summation rule is applied and $r^2=x^ix^i$ is the square of the distance of the observer from the central object. Note that in this notation $x^1=x$, $x^2=y$, $x^3=z$, $X^1=X$, $X^2=Y$ and $X^3=Z$. Since all quantities must be evaluated in the asymptotic limit, where $\frac{X^i}{r}\ll1$, one can write $1/R$ (as seen in Eq. (\ref{gammasol})) as 
\begin{eqnarray}
\label{invr}
\frac{1}{R}=\frac{1}{\sqrt{r^2-2x^iX^i+X^iX^i}}\approx\frac{1}{r}+\frac{x^iX^i}{r^3}~.
\end{eqnarray}

The energy-momentum tensor of a distribution of matter is defined as $_0T_{\mu\nu}=\rho c^2u_\mu u_\nu$, where $u_\mu=\eta_{\mu\nu}u^\nu$ is defined through Eq. (\ref{fourmomod}). For the case of a rotating body, which rotates around the $z$ axis in Cartesian coordinates, the components of $_0T_{\mu\nu}$ turn out to be
\begin{eqnarray}
\label{modenmom}
_0T_{00}=\rho \frac{m_\mathrm{R}}{m_\mathrm{I}}c^2,\,\,\,\,\,\,_0T_{01}=\rho\sqrt{\frac{m_\mathrm{R}}{m_\mathrm{I}}} cv\sin{\varphi}\equiv\sqrt{\frac{m_\mathrm{R}}{m_\mathrm{I}}}\,_{0}^0T_{01}\,\,\,\,\,\,\mathrm{and}\,\,\,\,\,\,_0T_{02}=-\rho\sqrt{\frac{m_\mathrm{R}}{m_\mathrm{I}}} cv\cos{\varphi}\equiv\sqrt{\frac{m_\mathrm{R}}{m_\mathrm{I}}}\,_{0}^0T_{02}~,%\,\,\,\,\,\,\mathrm{and\,\,all\,\,other}\,\,\,\,\,\,_0T_{\mu\nu}=0~,
\end{eqnarray}
where $\rho$ is the matter distribution density, $v$ is the rotational velocity of the central object, assumed to be small $v\ll c$, and $_{0}^0T_{01}$ and $_{0}^0T_{0}$ are the unmodified tensor components. Note that other tensor components vanish. Using the above energy-momentum tensor and Eq. (\ref{invr}), one can write the two components of the solution from Eq. (\ref{gammasol}) as
\begin{eqnarray}
\label{solsgamma}
\gamma_{01}=-\frac{\kappa}{2\pi}\frac{y}{r^3}\int Y\,_0T_{01}\,\mathrm{d}^3\mathbf{X}\,\,\,\,\,\,\mathrm{and}\,\,\,\,\,\,\gamma_{02}=-\frac{\kappa}{2\pi}\frac{x}{r^3}\int X\,_0T_{02}\,\mathrm{d}^3\mathbf{X}~,
\end{eqnarray}
where the axial symmetry of the system ensures that the $1/r$ terms and terms with other coordinate components vanish. 
Choosing the $z$-axis as the axis of rotation, 
%it is useful to write 
the special relativistic angular momentum reads
\begin{eqnarray}
\label{angmom}
J^z=\frac{1}{c}\int \left(X\,_0^0T^{02}-Y\,_0^0T^{01}\right)\,\mathrm{d}^3\mathbf{X}~.
\end{eqnarray}
The axial symmetry of the system also relates the two terms in the above definition as
\begin{eqnarray}
\label{axsymmangmom}
\int X\,_0^0T^{02}\,\mathrm{d}^3\mathbf{X}=-\int Y\,_0^0T^{01}\,\mathrm{d}^3\mathbf{X}=\frac{c}{2}J^z~.
\end{eqnarray}
Note that the energy-momentum tensor components in the definition of the angular momentum is contravariant. To make this consistent with the previous steps, it is straightforward to show the relations to the covariant energy-momentum tensor components $_0^0T^{01}=-\,_0^0T_{01}$ and $_0^0T^{02}=-\,_0^0T_{02}$. Using these relations, the second and third expressions in Eq. (\ref{modenmom}), and Eq. (\ref{axsymmangmom}), the solutions from Eq. (\ref{solsgamma}) become
\begin{eqnarray}
\label{gammaalmost}
\gamma_{01}=-\frac{\kappa\,c\,y}{4\pi r^3}\sqrt{\frac{m_\mathrm{R}}{m_\mathrm{I}}}J^z\,\,\,\,\,\,\mathrm{and}\,\,\,\,\,\,\gamma_{02}=\frac{\kappa\,c\,x}{4\pi r^3}\sqrt{\frac{m_\mathrm{R}}{m_\mathrm{I}}}J^z~.
\end{eqnarray}

What remains is to evaluate the constant $\kappa$. To do this, one must first consider the first integral of the geodesic equation 
\begin{eqnarray}
\label{firstint}
g_{\mu\nu}\frac{\mathrm{d}x^\mu}{\mathrm{d}s}\frac{\mathrm{d}x^\nu}{\mathrm{d}s}=1~.
\end{eqnarray}
%which relates its components.
%
To compare this result with the Newtonian limit, one neglects all terms $(1/r)^2$ and higher in the above expression, after plugging in the Kerr metric. One considers a geodesic trajectory for a particle orbiting the central object in the $x-y$ plane. After some algebraic manipulation and differentiating over $\varphi$ of Eq. (\ref{firstint}), %(see Ref. \cite{Papapetrou:1974gq}, page 72 for more details),
one obtains
\begin{eqnarray}
\label{geotonewt}
\frac{\mathrm{d}^2}{\mathrm{d}\varphi^2}\left(\frac{1}{r}\right)+\frac{1}{r}=-\frac{K}{2\ell^2}~,
\end{eqnarray}
where $\ell\equiv r^2\frac{\mathrm{d}\varphi}{\mathrm{d}s}$ is a constant of motion, related to angular momentum, for an orbiting particle. According to Newtonian mechanics, 
the energy of an orbiting particle is constant for any distance from the central object $r$, and is given by
\begin{eqnarray}
\label{newten}
\frac{m_\mathrm{I}\Dot{r}^2}{2}-\frac{Gm_\mathrm{G}M}{r}=\mathrm{const.}~,
\end{eqnarray}
where one can clearly differentiate between $m_\mathrm{I}$ and $m_\mathrm{G}$. Without loss of generality, one can assume that the particle is orbiting the central object in the $x-y$ plane. Then, one can write the particle position vector as $\mathbf{r}=r(\cos{\varphi},\sin{\varphi})$ and its angular momentum as $\mathbf{J}=\mathbf{r}\times\mathbf{p}=\sqrt{m_\mathrm{R}m_\mathrm{I}}\,c\,\ell\,\hat{z}$, where $\mathbf{p}$ is defined by the spatial components of Eq. (\ref{fourmomod}). Using the above, $\mathbf{J}=I\boldsymbol{\omega}=m_\mathrm{I}r^2\frac{\mathrm{d}\varphi}{\mathrm{d}t}\hat{z}$ and differentiating with respect to $\varphi$, one obtains
\begin{eqnarray}
\label{justnewt}
\frac{\mathrm{d}^2}{\mathrm{d}\varphi^2}\left(\frac{1}{r}\right)+\frac{1}{r}=\frac{m_\mathrm{G}}{m_\mathrm{R}}\frac{GM}{\ell^2c^2}~.
\end{eqnarray}
By comparing Eqs. (\ref{geotonewt}) and (\ref{justnewt}), one can identify the constant $K$, which is exactly the same as in the Schwarzschild and weak field cases (see Eq. (\ref{constK})). Note that the angular momentum, used to derive Eq. (\ref{justnewt}) corresponds to the angular momentum of a test particle orbiting the central object and is different from the angular momentum, defined in Eq. (\ref{angmom}), which corresponds to the rotation of the central object.

To obtain $\kappa$, one uses the first expression in Eq. (\ref{modenmom}) with Eq. (\ref{statlineinst}) to obtain
\begin{eqnarray}
\nabla^2\gamma_{00}=2\kappa\,\rho\,\frac{m_\mathrm{R}}{m_\mathrm{I}}c^2~,
\end{eqnarray}
which must correspond to the Poisson's equation for the gravitational potential in the weak field limit
\begin{eqnarray}
\nabla^2\phi=4\pi G\rho~.
\end{eqnarray}
Note that the above Poisson's equation does not obtain corrections, since the $m_\mathrm{G}/m_\mathrm{I}$ ratio would modify it on both sides equally, which would then cancel. By comparing the above two equations, it turns out that
\begin{eqnarray}
\gamma_{00}=\frac{m_\mathrm{R}}{m_\mathrm{I}}\frac{\kappa\,c^2}{2\pi G}\phi~,
\end{eqnarray}
and by using the inverse relation of Eq. (\ref{gammah}), the $g_{00}$ component in the weak field approximation turns out to be
\begin{eqnarray}
g_{00}=1+\frac{m_\mathrm{R}}{m_\mathrm{I}}\frac{\kappa\,c^2}{4\pi G}\phi=1+\frac{K}{r}~.
\end{eqnarray}
Precisely the same terms appear in components $g_{11}$, $g_{22}$ and $g_{33}$. Since the constant $K$ and potential $\phi$ are known (see Eq. (\ref{constK})), the modified constant $\kappa$ can easily be identified as
\begin{eqnarray}
\label{einconst}
\kappa=\frac{m_\mathrm{G}m_\mathrm{I}}{m_\mathrm{R}^2}\frac{8\pi G}{c^4}~,
\end{eqnarray}
thus effectively modifying the Einstein equations as
\begin{eqnarray}
R_{\mu\nu}-\frac{1}{2}Rg_{\mu\nu}=\frac{m_\mathrm{G}m_\mathrm{I}}{m_\mathrm{R}^2}\frac{8\pi G}{c^4}T_{\mu\nu}~.
\end{eqnarray}

By plugging the modified Einstein constant from Eq. (\ref{einconst}) in the expressions from Eq. (\ref{gammaalmost}) and using Eq. (\ref{gammah}), one obtains
\begin{eqnarray}
h_{01}=\gamma_{01}=-\frac{m_\mathrm{G}m_\mathrm{I}^{1/2}}{m_\mathrm{R}^{3/2}}\frac{2\,G}{c^3}\frac{y}{r^3}J^z\,\,\,\,\,\,\mathrm{and}\,\,\,\,\,\,h_{02}=\gamma_{02}=\frac{m_\mathrm{G}m_\mathrm{I}^{1/2}}{m_\mathrm{R}^{3/2}}\frac{2\,G}{c^3}\frac{x}{r^3}J^z~.
\end{eqnarray}
At this point one can define the constant $a\equiv\frac{J^z}{Mc}$. Using $a$ and the above information, one can write the line element of the metric of a rotating object as
\begin{eqnarray}
\label{asymlim}
\mathrm{d}s^2&=&\left(1-\frac{m_\mathrm{G}}{m_\mathrm{R}}\frac{2GM}{c^2r}\right)c^2\mathrm{d}t^2-\left(1-\frac{m_\mathrm{G}}{m_\mathrm{R}}\frac{2GM}{c^2}\right)(\mathrm{d}x^2+\mathrm{d}y^2+\mathrm{d}z^2) \nonumber \\
&-&\frac{m_\mathrm{G}m_\mathrm{I}^{1/2}}{m_\mathrm{R}^{3/2}}\frac{4GM}{c^2}\frac{y}{r^3}a\,\,c\,\mathrm{d}t\,\,\mathrm{d}x+\frac{m_\mathrm{G}m_\mathrm{I}^{1/2}}{m_\mathrm{R}^{3/2}}\frac{4GM}{c^2}\frac{x}{r^3}a\,\,c\,\mathrm{d}t\,\,\mathrm{d}y~.
\end{eqnarray}
Comparing the above line element with the asymptotic limit when $r\longrightarrow\infty$ of the line element from Eq. (\ref{kerrcart}), one can identify the remaining constant $L$ as
\begin{eqnarray}
L=\sqrt{\frac{m_\mathrm{I}}{m_\mathrm{R}}}a~.
\end{eqnarray}
The obtained constants $K$ and $L$ are then plugged in the metric from Eq. (\ref{kerrconst}), which is then inverted to obtain the contravariant Kerr metric as \begin{eqnarray}
\label{kerrmod}
g^{\mu\nu}(m_\alpha)=\left[\begin{array}{cccc}
 \frac{\Sigma^2(m_\alpha)}{\rho^2(m_\alpha)\Delta(m_\alpha)}    & 0 & 0& 2\frac{m_\mathrm{G}m_\mathrm{I}^{1/2}}{m_\mathrm{R}^{3/2}}\frac{GMar}{c^2\rho^2(m_\alpha)\Delta(m_\alpha)} \\
0 & -\frac{\Delta(m_\alpha)}{\rho^2(m_\alpha)} & 0 & 0 \\
0 & 0 & -\frac{1}{\rho^2(m_\alpha)} & 0 \\
2\frac{m_\mathrm{G}m_\mathrm{I}^{1/2}}{m_\mathrm{R}^{3/2}}\frac{GMar}{c^2\rho^2(m_\alpha)\Delta(m_\alpha)} & 0 & 0 & -\frac{\Delta(m_\alpha)-\frac{m_\mathrm{I}}{m_\mathrm{R}}a^2\sin^2{\theta}}{\rho^2(m_\alpha)\Delta(m_\alpha)\sin^2{\theta}}
\end{array}\right]~,
\end{eqnarray}
where $\Sigma^2(m_\alpha)=\big(r^2+\frac{m_\mathrm{I}}{m_\mathrm{R}}a^2\big)^2\!\!-\frac{m_\mathrm{I}}{m_\mathrm{R}}a^2\Delta\sin^2{\theta}$, $\,\,\Delta(m_\alpha)=r^2-\frac{m_\mathrm{G}}{m_\mathrm{R}}\frac{2GM}{c^2}r+\frac{m_\mathrm{I}}{m_\mathrm{R}}a^2$ and $\rho^2=r^2+\frac{m_\mathrm{I}}{m_\mathrm{R}}a^2\cos^2{\theta}$.

\subsection*{Feshbach-Villars Formalism in Curved Space-time}\setcurrentname{Feshbach-Villars Formalism in Curved Space-time}
\label{appP}

For convenience the Feshbach-Villars formalism is used for bosons,  
suitably generalized to background curved space-times. 
The effective Hamiltonian for bosons in a curved space-time background in the Feshbach-Villars formalism takes the form
%%%
%\footnote{While we have developed the Hamiltonian in the given representation independently, there has been earlier authors who worked on a similar formalism using different representations \cite{Cognola:1985qg,Tagirov:1999nc}.}.
%%%
\begin{eqnarray}
\label{hamiltonian}
\hat{H}=\tau_3 \frac{1}{\sqrt{g^{00}}}mc^2-(\tau_3+i\tau_2)\frac{g^{ij}}{\sqrt{g^{00}}}\frac{\hat{p}_i\hat{p}_j}{2m}+(\tau_3+i\tau_2)\frac{g^{0i}}{\sqrt{g^{00}}}\frac{\hat{p}_0\hat{p}_i}{m}~,
\end{eqnarray}
where $\tau_k$ ($k=1,2,3$) are the Pauli matrices, $g^{\mu\nu}$ is the metric of the curved space-time background ($\mu,\nu=0,1,2,3$ with $0$ representing the time component and $i,j=1,2,3$ the spatial components), $m$ is the mass of the boson, $c$ the speed of light, $\hat{p}_0=i\frac{\hbar}{c}\nabla_0$, $\hat{p}_i=-i\hbar\nabla_i$, $\hbar$ the reduced Planck constant and $\nabla_\mu$ the covariant derivative compatible with the background metric. Note that the Feshbach-Villars formalism in curved space-time has been previously studied in other  representations, as seen in Refs. \cite{Cognola:1985qg,Tagirov:1999nc}. However, the formulation with the simple representation given in Eq. (\ref{hamiltonian}) is introduced 
%for the first time to the best of one's knowledge, 
in this work. Also, the Pauli matrices in this formalism are %simply used for convenience, 
used here simply for convenience and are in no way related to 
%to formulate a Hamiltonian in such a representation and are not related to 
the spin of the particle.
They are also not affected by the 
curvature of space-time, unlike for the Dirac matrices, 
%discussed in the next section, 
which are relevant for particles with half-integer spins.

The eigenstates of the effective Hamiltonian in Eq. (\ref{hamiltonian}) are represented by a pair of scalar functions $\varphi$ and $\chi$ as
\begin{eqnarray}
\label{wfunc}
\Psi=\left[\begin{array}{c}
\varphi  \\
\chi
\end{array}\right]~.
\end{eqnarray}
The above definition of scalar functions $\varphi$ and $\chi$ is used for convenience to formulate a Hamiltonian corresponding to relativistic bosons. They are related to the two degrees of freedom of the wavefunction from the Klein-Gordon equation, which correspond to the two possible charge states \cite{Feshbach:1958wv}. The Hamiltonian in Eq. (\ref{hamiltonian}) and the wavefunction in Eq. (\ref{wfunc}) satisfy a Schr\"{o}dinger-like equation
\begin{eqnarray}
\label{schrodi}
\hat{H}\Psi=i\hbar\,\partial_0\Psi~.
\end{eqnarray}
In the above, the partial derivative with respect to time is identical to the 
covariant derivative, i.e., $\partial_0\Psi=\nabla_0\Psi$, since $\Psi$ consists of scalar functions. This identity is taken into account in the following steps.

%on a scalar function is exactly the same as the covariant time derivative $\nabla_0\Psi=\partial_0\Psi$.

The wavefunction from the Klein-Gordon equation in this formalism is defined as $\Phi_{_{\!\mathrm{KG}}}=\varphi+\chi$. To see 
that the above formalism indeed represents the Klein-Gordon equation in curved space-time, one plugs Eqs. (\ref{hamiltonian}) and  (\ref{wfunc}) in Eq. (\ref{schrodi}) to obtain two coupled differential equations
\begin{eqnarray}
\label{coup1}
g^{ij}\frac{\hbar^2\nabla_i\nabla_j}{2m}\Phi_{_{\!\mathrm{KG}}}+g^{0i}\frac{\hbar^2\nabla_0\nabla_i}{mc}\Phi_{_{\!\mathrm{KG}}} +mc^2\varphi=i\hbar\sqrt{g^{00}}\nabla_0\varphi~,
\end{eqnarray}
and
\begin{eqnarray}
\label{coup2}
-g^{ij}\frac{\hbar^2\nabla_i\nabla_j}{2m}\Phi_{_{\!\mathrm{KG}}}-g^{0i}\frac{\hbar^2\nabla_0\nabla_i}{mc}\Phi_{_{\!\mathrm{KG}}} -mc^2\chi=i\hbar\sqrt{g^{00}}\nabla_0\chi~.
\end{eqnarray}
The above pair of equations represent a coupled system of equations for $\varphi$ and $\chi$. However, it is required to obtain the equation of motion for $\Phi_{_{\!\mathrm{KG}}}$, since it is defined as the wavefunction from the Klein-Gordon equation. To achieve this, Eqs. (\ref{coup1}) and (\ref{coup2}) can be used to retrieve additional relations between $\varphi$ and $\chi$. Eqs. (\ref{coup1}) and (\ref{coup2}) may then be added together and derived over time to obtain a useful identity
\begin{eqnarray}
\label{iden1}
mc^2(\varphi-\chi)=i\hbar\sqrt{g^{00}}\nabla_0\Phi_{_{\!\mathrm{KG}}}\,\,\,\,\,\,\longrightarrow\,\,\,\,\,\,mc^2(\nabla_0\varphi-\nabla_0\chi)=i\hbar\sqrt{g^{00}}\nabla_0\nabla_0\Phi_{_{\!\mathrm{KG}}}~.
\end{eqnarray}
In the above, the identity $\nabla_\alpha g^{\mu\nu}=0$ has been used. Furthermore, subtracting Eq.(\ref{coup2}) from (\ref{coup1}), one obtains
\begin{eqnarray}
\label{iden2}
g^{ij}\frac{\hbar^2\nabla_i\nabla_j}{m}\Phi_{_{\!\mathrm{KG}}}+2\,g^{0i}\frac{\hbar^2\nabla_0\nabla_i}{mc}\Phi_{_{\!\mathrm{KG}}} +mc^2\Phi_{_{\!\mathrm{KG}}}=i\hbar\sqrt{g^{00}}(\nabla_0\varphi-\nabla_0\chi)~.
\end{eqnarray}
Plugging Eq. (\ref{iden1}) in Eq. (\ref{iden2}) and with a bit of algebraic manipulation, one obtains the Klein-Gordon equation in curved space-time \cite{qftcst}
\begin{eqnarray}
\left(g^{\mu\nu}\nabla_\mu\nabla_\nu+\frac{m^2c^2}{\hbar^2}\right)\Phi_{_{\!\mathrm{KG}}}=0~.
\end{eqnarray}
As expected, in the non-relativistic limit and in flat space-time, the standard quantum mechanical results are obtained. 
This shows that Eq. (\ref{hamiltonian}) is 
indeed the right Hamiltonian for a bosonic particle in an arbitrary gravitational field. This demonstrates that  the Klein-Gordon equation, 
relevant for spin $0$ particles can be written as an effective Hamiltonian
for a Schr\"odinger-like equation. 

By implementing the different masses to the Hamiltonian in Eq. (\ref{hamiltonian}), as seen in the main paper, and following the above procedure, one obtains a modified Klein-Gordon Equation in curved space-time
\begin{eqnarray}
\left(g^{\mu\nu}(m_\alpha)\tilde{\nabla}_\mu\tilde{\nabla}_\nu+\frac{m_\mathrm{R}^2c^2}{\hbar^2}\right)\Phi_{_{\!\mathrm{KG}}}=0~,
\end{eqnarray}
where $\tilde{\nabla}_\mu=\left(\frac{1}{c}\nabla_0,\sqrt{\frac{m_\mathrm{R}}{m_\mathrm{I}}}\boldsymbol{\nabla}\right)$, which is consistent with the quantum version of Eq. (\ref{fourmom}), when multiplied by $i\hbar$.

\end{document}